\documentclass[12pt]{article}

\usepackage{cite}
\usepackage{epsfig}
\usepackage{graphics}
\usepackage{inputenc}
\inputencoding{latin1}

\begin{document}

\sloppy

\begin{titlepage}

{\par\raggedleft 
 \texttt{UCL/HEP 2002-04}\\
\texttt{October} \texttt{2002}\par}
\bigskip{}
                                                                                                   
\bigskip{}
{\par\centering \textbf{\large JetWeb: A WWW Interface and Database for Monte Carlo Tuning and Validation}
\large \par}
\bigskip{}

{\par\centering J. M. Butterworth \\
S. Butterworth\\
\par}
\bigskip{}

{\par\centering {\small Department of Physics \& Astronomy}\\
{\small University College London}\\
{\small Gower St. London WC1E 6BT} \\
{\small England}\small \par}

\bigskip{}

\begin{abstract}
\noindent

A World Wide Web interface to a Monte Carlo validation and tuning facility is
described.  The aim of the package is to allow rapid and reproducible
comparisons to be made between detailed measurements at high-energy
physics colliders and general physics simulation packages.  The
package includes a relational database, a Java servlet query and
display facility, and clean interfaces to simulation packages and their
parameters.

\end{abstract}

{\it PACS:} 

\end{titlepage}

\section{Introduction}
\label{sec:intro}

We present a web-based facility for the comparison of measurements of
hadronic final states and the expectations of physics calculations and
simulations. The physics motivation is discussed. Then follows an
outline of the design and technology used in the JetWeb
server. Section~4 contains a user guide to the server, as presented on
the web, while Sections ~5 and ~6 cover the implementation of the
server with a view to further development and future plans.

\section{The physics problem}

Particle physics experiments at high-energy accelerators have provided
a wealth of data on the final state in electron-positron,
lepton-proton and proton-antiproton interactions. These data have seen
the triumph of the standard model in precision electroweak
measurements and the verification of the QCD sector of the standard
model to a reasonable degree of precision. 

Despite these successes, the final state in hadron-hadron collisions
is in general poorly understood. The hadronisation stage of a
collision is not calculable in perturbative QCD, and calculations of
(for example) multijet production are impractical at higher orders in
QCD. Other problems include the description of non-perturbative
remnant-remnant interactions, and the determination of the intrinsic
transverse momentum of partons in the incoming hadron. All these areas
are calculated and/or modelled to some degree in general purpose
Monte-Carlo simulation programs such as Pythia~\cite{pythia} and
Herwig~\cite{herwig}. These packages provide an invaluable tool in
several distinct ways. 
\begin{itemize}

\item They provide physically motivated input to detector simulations,
thus facilitating the evaluation of acceptances and resolutions for
detectors.

\item They allow the estimation of the effects of hadronisation and
other uncalculated effects on cross sections so that physically
observable hadronic cross sections may be compared to partonic
variables calculated to high orders in QCD.

\item They allow predictions to be made for collisions at future
colliders, which are vital input to the motivation and design of
facilities.

\end{itemize}

For all these reasons, consistent tuning of the free parameters of
these generators, and confirmation of the physics assumptions they
contain, is of critical importance for the accuracy and interpretation
of measurements at current and future high-energy colliders.  However,
determining whether such models are consistent with the wide range of
available data is a non-trivial matter since the measurements are made
with a variety of colliding beams, in many different regions of phase
space, and for many complex observables. Detailed and successful fits
have been performed using LEP data but these give no
information on physics specific to incoming hadrons. In addition,
there is an urgent need for the preservation of expertise and access
to this data.

The JetWeb server is designed to provide the ability to compare
quickly and efficiently existing or future calculations and models
with all relevant collider data. This is an ambitious and long term
(indeed indefinite) task. Currently a subset of HERA, LEP and Tevatron
jet data are included and only the Herwig and Pythia generators are
used. However, the server is designed to be scalable, with clean
interfaces between the various technologies used for ease of
development and maintenance and the inclusion of new data and
simulations.

\section{The JetWeb server}

\begin{figure}
\psfig{file=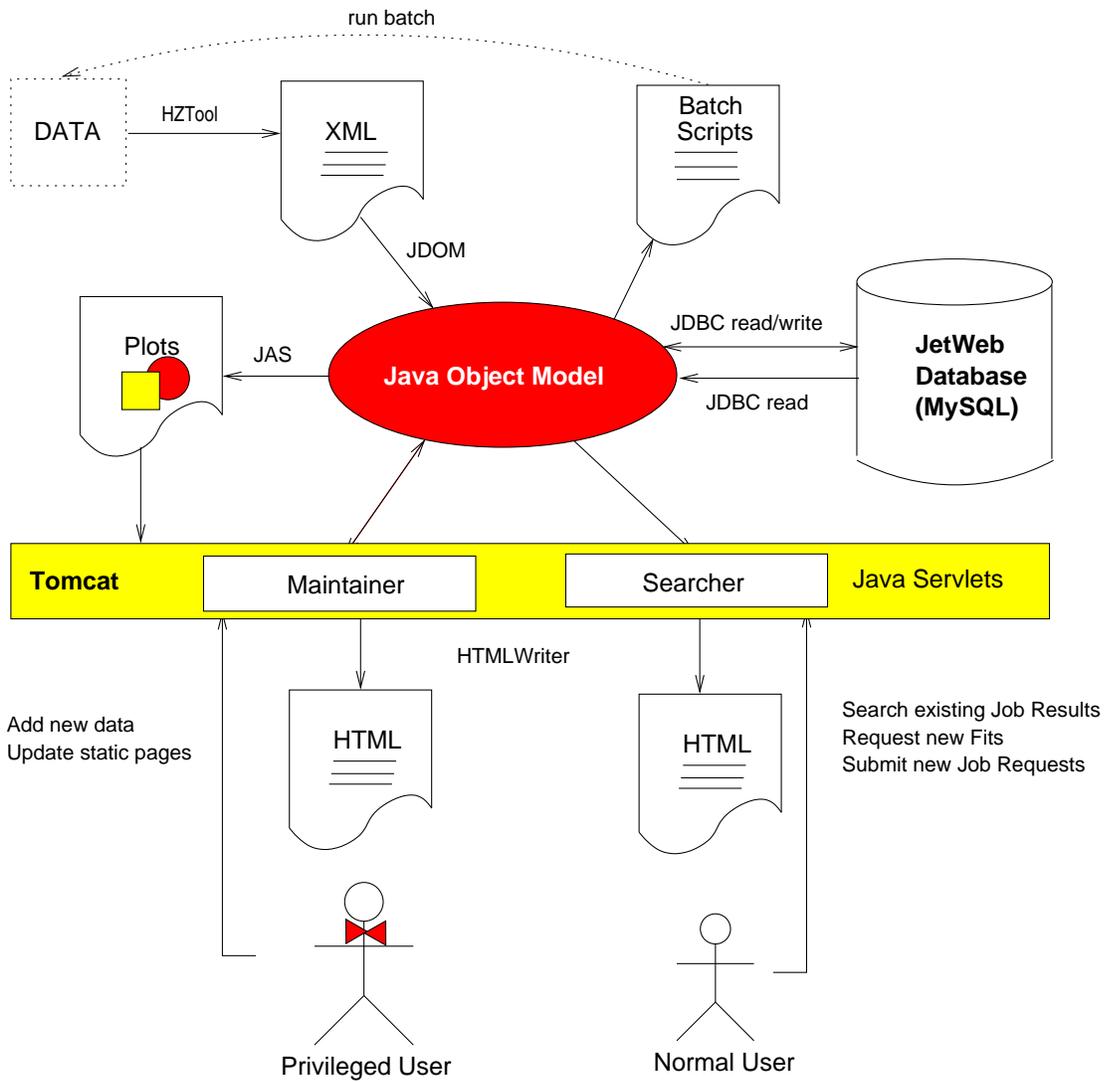,width=\textwidth}
\caption{\label{fig:jwfig} Components of the JetWeb server.}
\end{figure}

The components of the JetWeb server and their functions and interactions
are illustrated in Figure~\ref{fig:jwfig}.

\begin{figure}
\psfig{file=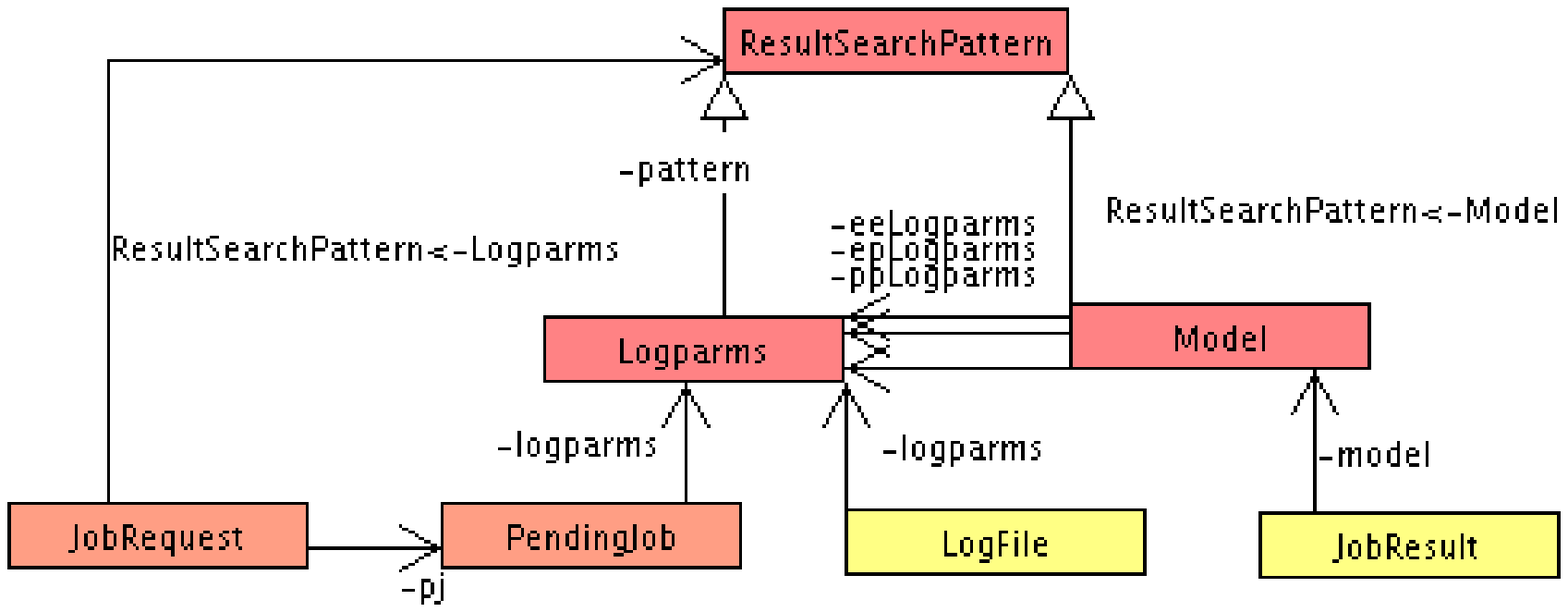,width=\textwidth}
\caption{\label{fig:jw1fig} {\bf JetWeb Object Model:
ResultSearchPattern.}  {\it ResultSearchPattern} is a container for a
set of parameters specifying a database search. {\it Model} and {\it
Logparms} are both specialised types of {\it ResultSearchPattern}. A
{\it Model} specifies a complete set of parameters including a unique
generator but not a collider, while a {\it Logparms} defines a
unique collider along with the subset of parameters it requires. {\it
JobRequest, PendingJob} and {\it Logfile} all describe a single batch
job with parameters specifed by a {\it Logparms}. {\it JobRequest} is
constructed from a {\it ResultSearchPattern}; that is, the {\it
ResultSearchPattern} parameters specify the job to be run. {\it
PendingJob} is built when a {\it JobRequest} is submitted and exists
to provide status information about unfinished batch jobs. {\it
Logfile} is built from the output log file of a batch job.  A {\it
JobResult} is constructed from a Model and a set of data to which it
is compared.}
\end{figure}

\begin{figure}
\psfig{file=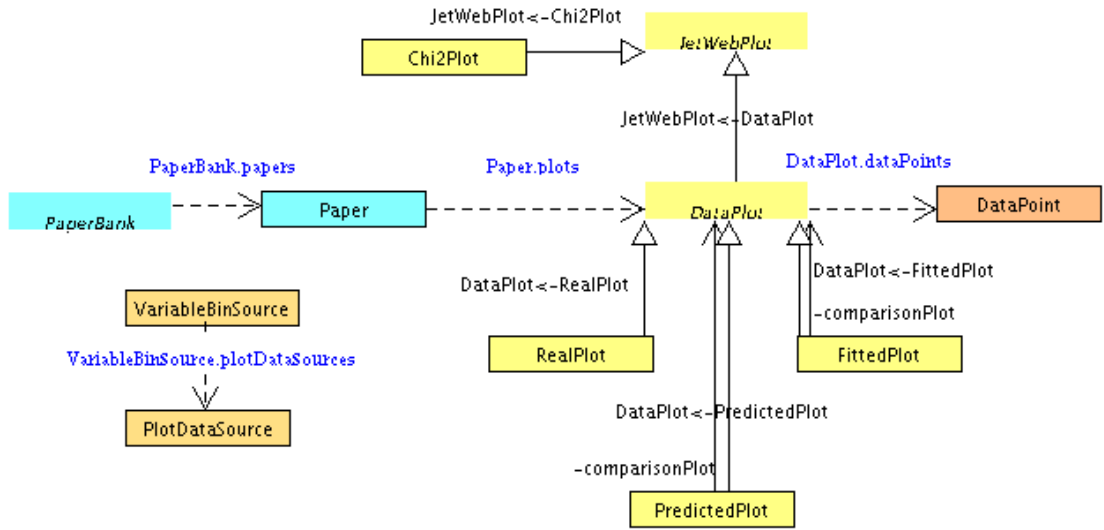,width=\textwidth}
\caption{\label{fig:jw2fig} {\bf JetWeb Object Model: Papers and
Plots.} Data is held in the object model in this series of paper and
plot objects. At the top level, {\it PaperBank} is a container for all
the {\it papers}. Each {\it Paper} contains a collection of {\it
plots}. {\it JetWebPlot} is the fundamental plot class and
incorporates JAS functionality to output gifs and also includes the
plotML read/write methods. There are three types of {\it DataPlot};
{\it RealPlot} (real measured data from experiments), {\it
PredictedPlot} (MC data) and {\it FittedPlot} (MC data that has been
compared to real data). {\it Chi2Plot} is not a type of {\it
DataPlot}. It plots $\chi^2$ scatter for a selection of fits. Each
{\it DataPlot} contains a set of {\it DataPoints}. {\it
VariableBinSource} is a container for a set of plot data sources to
handle variable bin histograms. {\it PlotDataSource} is the JetWeb 1D
Histogram object, and is the jetweb implementation of the JAS interface
{\it Rebinnable1DHistogramData}. A {\it PlotDataSource} can be constructed from a
set of {\it DataPoints} allowing a histogram to be written from a {\it DataPlot}. NB, classes represented by unbounded boxes are Abstract.}
\end{figure}

\begin{figure}
\psfig{file=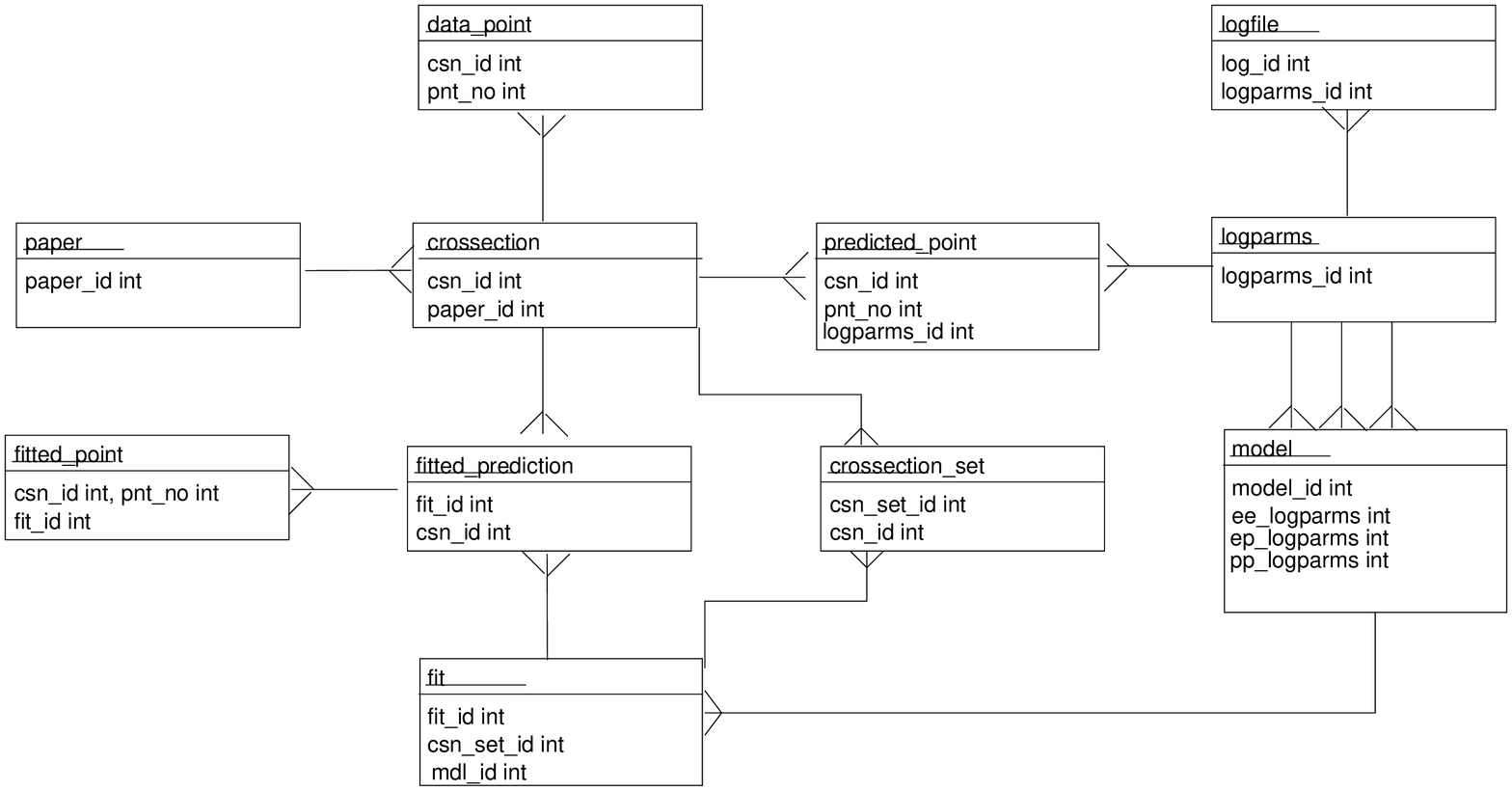,width=\textwidth}
\caption{\label{fig:jwdbfig} {\bf JetWeb Database.} The diagram shows
the tables of the JetWeb database, their primary keys, the
relationships between the tables and the fields on the tables
pertaining to those relationships.  A {\it paper} (identified by
paper\_id) may have one or more {\it crosssections}, while each {\it
crosssection} record (identified by a csn\_id) has a single paper\_id
associated with it. A {\it crosssection} (equivalent to {\it Plot} in the
java object model) may have a set of {\it data\_points} or {\it
predicted\_points} or a {\it fitted\_prediction}.  A {\it fit}
compares a {\it model} to a set of {\it crosssections}, as specified
by its {\it crosssection\_set}.  Each {\it model} may have up to three
associated logparms records; one each for electron-electron,
electron-proton and proton-proton.}
\end{figure}

At the centre of JetWeb is a java object model (Figures
~\ref{fig:jw1fig} and ~\ref{fig:jw2fig}) containing the properties
and interactions of Models, Papers, Plots and Fits. The data
underlying this model are stored in the JetWeb database -  a
MySQL database (Figure~\ref{fig:jwdbfig}). The approximate
correspondences between tables and object model classes are given in Table~1.

\begin{table}
\begin{center}
\begin{tabular}{|c|c|}  \hline
Class                & Table \\ \hline\hline
Logparms             & logparms     \\
                     & herwig\_parameters \\ 
                     & pythia\_parameters \\ \hline
Model                & model        \\ 
                     & herwig\_model \\ 
                     & pythia\_model \\ \hline
JobResult            & fit         \\ \hline
Paper                & paper        \\ \hline
RealPlot             & crossection  \\ 
DataPoint            & data\_point   \\ \hline
PredictedPlot        & crossection  \\ 
DataPoint            & predicted\_point    \\ \hline
FittedPlot           & fitted\_prediction  \\ 
DataPoint            & fitted\_point       \\ \hline
LogFile              & logfile  \\ 
PlotSelection        & crosssection\_set        \\ \hline 
\end{tabular}
\caption{ Approximate correspondence  
between classes in the object model
and tables in the database.}
\end{center}
\end{table}

A ``Model'' completely specifies a unique generator, version and set
of parameters. A ``Logparms'' consists of a subset of these parameters
relevant to a specific set of colliding beams (for example,
proton-proton logparms do not define a photon parton distribution, and
photon-photon logparms likewise do not specify a proton parton
distribution). Consquently one model can currently have up to three
logparms associated with it ($ee$, $ep$ and $pp$). A ``Logfile'' for an individual simulation run is associated with a single logparms.  A
``Paper'' encapsulates the measured data from a single publication and
is associated with measured cross sections.  A ``Fit'' contains the
results of a comparison between real data and the predictions of a
specific model. For more detail on the constituents of the object
model, see  Figures~\ref{fig:jw1fig}
and ~\ref{fig:jw2fig}.

Data is added to the database as follows:
\begin{itemize}
\item MC results from HZTOOL~\cite{hztool} are processed by a FORTRAN
routine which reads in HBOOK~\cite{hbook} histogram files and writes
out histograms in the form of XML files in the Java Analysis Studio~\cite{JAS}
(JAS) plotML DTD format~\cite{plotML}.
\item The JetWeb XMLReader facility is a Java class, used to convert
plotML data into the Java object model. It processes the XML data
using the JDOM API~\cite{jdom}, wrapping a SAX parser.
\item Data held within the Java object model is written to the JetWeb
database via JDBC~\cite{jdbcdriver}. The interface between object
model and database is provided by the database manager classes within
the ucl.hep.jetweb.db package.
\end{itemize}
These stages are initiated by commands from the Maintainer (privileged
access user interface, see below).  The user accesses JetWeb via the
web user interface (UI), which consists of Java servlets run on a
Tomcat~\cite{Tomcat} server, delivering HTML pages written using the
JetWeb HTMLWriter facility (ucl.hep.jetweb.html package). The servlets
access the JetWeb database via the JDBC calls encapsulated within the
java object model.

There are two levels of access to JetWeb:
\begin{itemize}
\item Normal access allows the user to query the JetWeb database for
existing Job Results. The user can fit these results to the data from a
range of papers stored on the database. 

Fit data is loaded from the database into the object model. If this
fit has been requested before, the static webpages will already
exist. A comparison is made between the timestamp in the database and
the date of these pages. If the pages are more up-to-date, they are
served to the user (This avoids the extremely time consuming
regeneration of histogram graphics files).  If the pages predate any
relevant data in the database they are regenerated.  In this case an
HTML summary of the fit is written, and ucl.hep.jetweb.plots.DataPlot
and its subclasses access JAS to write gif plot files. Plot
files are incorporated into HTML generated by servlets using the
HTMLWriter utility. The graphics generation is done in a separate
thread to improve response time for the user.

If there are no results stored for the user's specified set of
parameters, a new job request with the required parameters can be
generated. The job requests are moderated and submitted by an
administrator on the server side.

\item Privileged access is available to allow maintenance of the
JetWeb Server. It allows the update of the static webpages associated
with the application. It controls the upload of new data to the JetWeb
database. It also manages the running of new batch jobs, facilitating
submission of new jobs and monitoring of incomplete runs.
\end{itemize}

\section{User guide}
This section outlines the expected use of the web pages
(http://jetweb.hep.ucl.ac.uk) for a normal (unprivileged) user.

\begin{figure}
\psfig{file=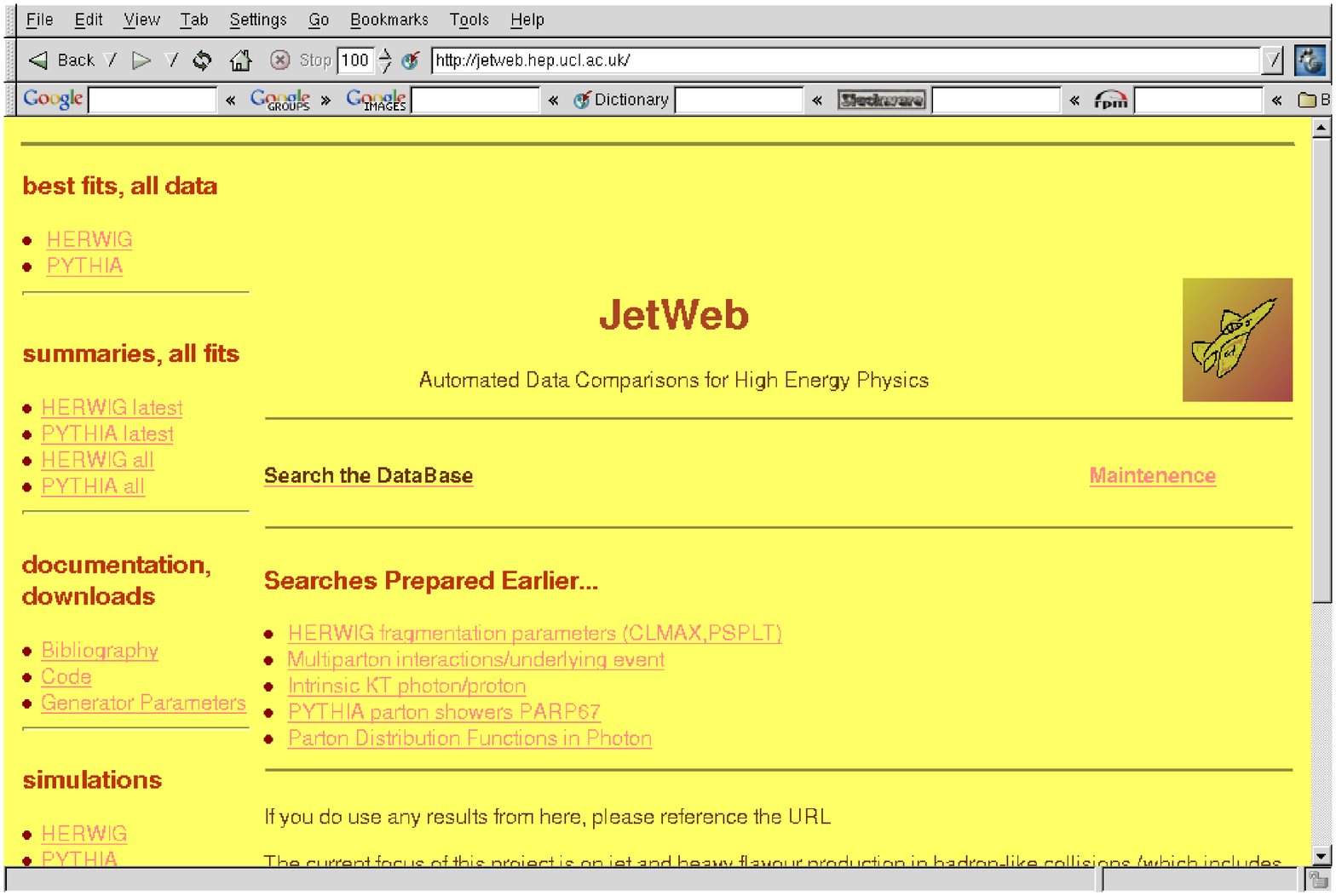,width=\textwidth}
\caption{\label{fig:frontpagefig} JetWeb Application Welcome page.}
\end{figure}

The top-level welcome page (Figure~\ref{fig:frontpagefig}) provides the user with several options.  Down the left hand side are various
short cuts to static web pages. These include
\begin{itemize}
\item
A short-cut to the current best fit for each simulation package used.
\item
Lists of all fits currently in the database.
\item
An extensive bibliography with references to the data used in JetWeb, presentations
using results from JetWeb, and other related work.
\item
Documentation of the simulations currently available in HZTOOL.
\item
Links to the experiments whose data are used in JetWeb.
\end{itemize}

In the central section of the page are links allowing the user to
\begin{itemize}
\item Query the database directly
\item Access previously prepared searches
\end{itemize}

\subsection{Querying the database for existing results}

\subsubsection{Prepared searches}

A selection of standard queries relating to specific
parameters or physics studies are listed on the JetWeb welcome
page. Following the links from here gives access to all the fits as
well as a summary of conclusions to be drawn from them.

\subsubsection{Interactive searches}
\begin{figure}
\psfig{file=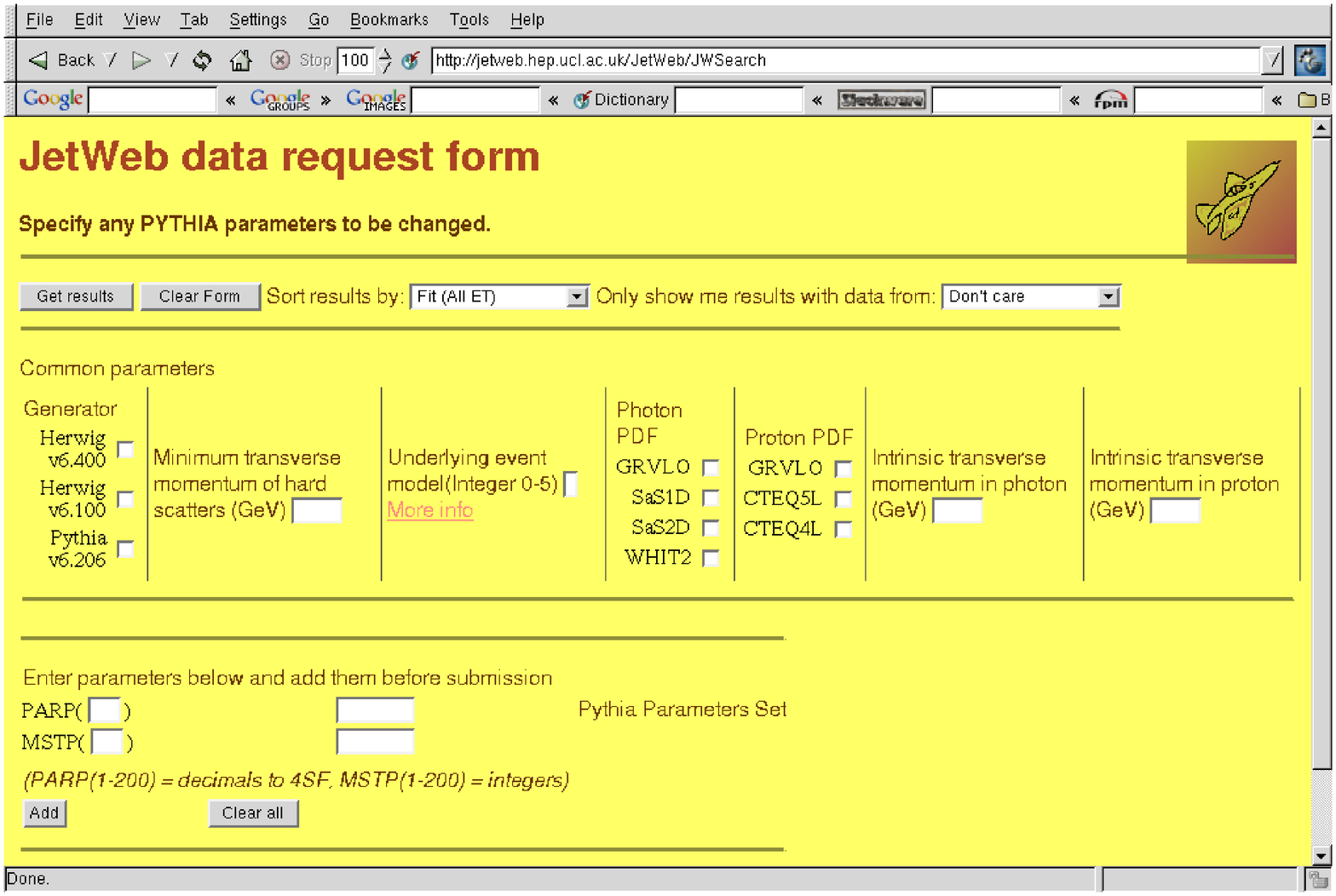,width=\textwidth}
\caption{\label{fig:searchpagefig} JetWeb Search Page.}
\end{figure}

On the welcome page, clicking `Search the Database' causes a parameter
input form to be displayed, as shown in Figure ~\ref{fig:searchpagefig}.
\begin{itemize}
\item Common parameters; various parameters common to all generators
can be entered on first part of form
\item Generator specific parameters: The search may be restricted to
specialised parameters for a particular generator. Currenly HERWIG and
PYTHIA are implemented. 
\begin{itemize}
\item HERWIG parameters. Check Generator - HERWIG. Click `Change
HERWIG parameters'. This will display additional fields for the HERWIG
parameters to be entered.
\item PYTHIA parameters. Check Generator - PYTHIA. Click `Change
PYTHIA parameters'. Boxes will be displayed to enter the number and
value of each MSTP and PARP parameter to be specified. Enter the values
and press `Add'. MSTP and PARP parameters which have been set will be
displayed on the form. To remove a value, enter its number in the
relevant field, but leave the corresponding box blank, then press
`Add'. Specify all the parameters required in this way before
continuing with the search.
\end{itemize}
\item The sort order of results can be specified by choosing a value from the
`sort results...' list at the top of the form.
\item Results can be filtered by selecting a value for the `only show
results..' list at the top of the form. Some models in the
database will have been compared to a subset of the available
data only. This option allows the user to exclude any models that have not
been compared to the chosen data.
\item Submit a search by clicking `Get Results' at the top of the form.
\end{itemize}

\subsubsection{Query output}
If a search is successful, brief summaries of the
available results will be listed. Each Job Result will also have a
link (`Plots etc') for access to a detailed description of the Fit and
the actual histograms showing the comparison (Figure~7).

If there are no data corresponding to the input search parameters, the JetWeb
Request form will be displayed with the required parameters filled in. From
here, a request can be made for a new job to be run.

\subsection{Request new jobs}
\begin{figure}
\psfig{file=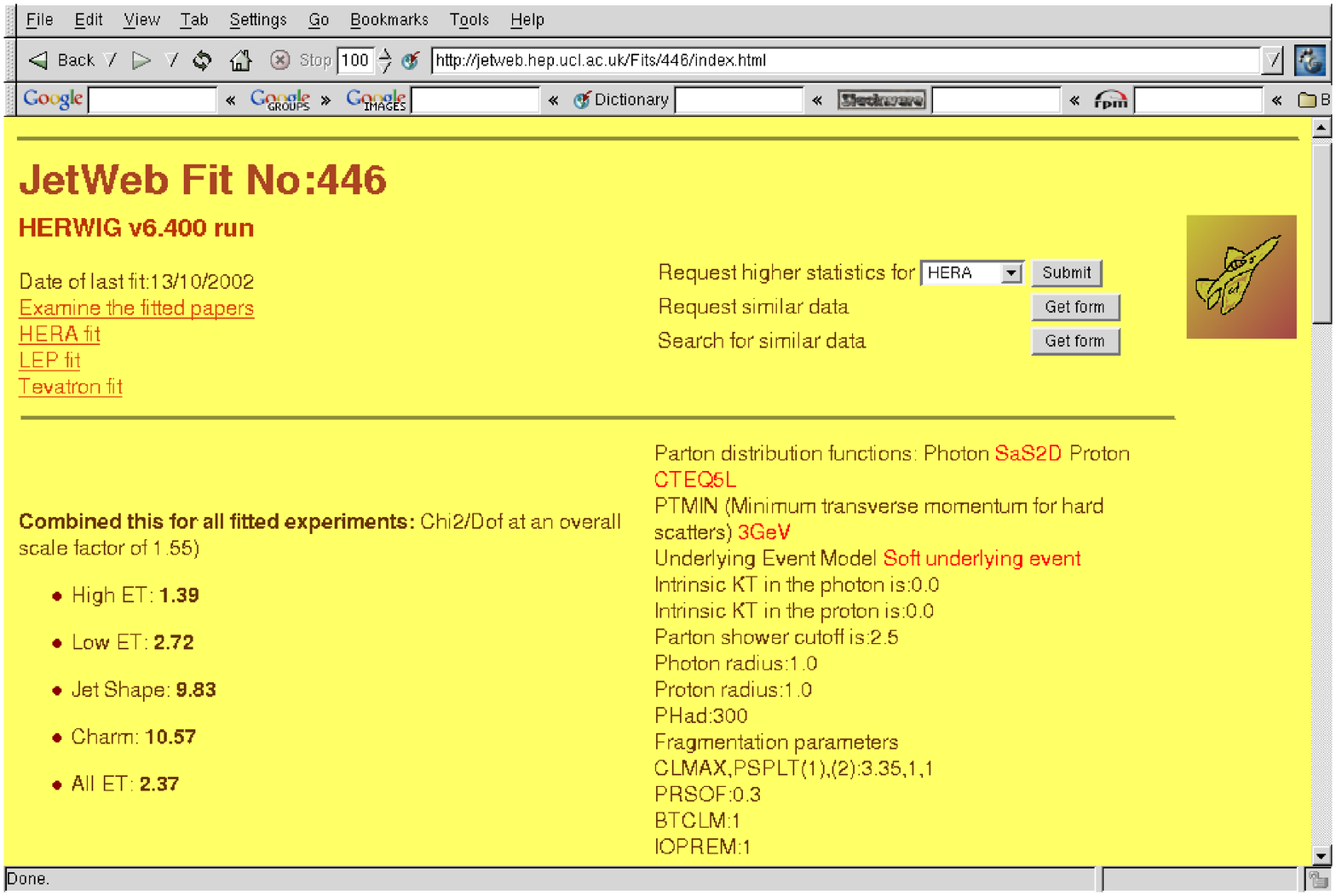,width=\textwidth}
\caption{\label{fig:fitpagefig} Fit Details Page.}
\end{figure}

Job requests can be made in two contexts
\begin{itemize}
\item Search fails to find results - this brings up the JobRequest
form with the failed search parameters filled in. 
\item A request for additional data can be made from the Fit Details
page, Figure~\ref{fig:fitpagefig}, for each job.  First, the collider must be selected. More data for a particular collider (with unchanged job parameters) can be requested using the `Request highwer statistics' option, while `Request similar data' allows a new job to be specified with altered parameters. The latter option accesses the job request form, allowing the user to vary a subset
of parameters while guaranteeing that all others will remain identical to
those of the originally selected fit.
\end{itemize}

\section{Developer guide}
This section is intended to outline ways in which the application
could be altered or enhanced to expand its scope. It should be used as a guide for developers in conjunction with JavaDoc for the
JetWeb java application, and other documentation on the JetWeb website
(jetweb.hep.ucl.ac.uk).

\subsection{To switch database}
MySQL is a simple relational SQL database with a JDBC driver
available.  The database is accessed from the java object model using
the ucl.hep.jetweb.db package. Minimal changes would be required to
convert to another SQL database (probably just changing the relevant
JDBC driver). For conversion to a different type of database, new
database manager classes would be required, implementing the same
interface as the existing DBManager, DBFitManager and DBPlotManager.

\subsection{To switch user interface}
The interface of the ucl.hep.jetweb.servlet package would have to be
implemented by any alternative UI. This application was originally run
using a Java Swing UI. It should be very straightforward to implement
a simple command-line interface. A new UI would replace the ucl.hep.jetweb.servlet package; accessing the same methods on ucl.hep.jetweb.ResultManager.

\subsection{To add new measurements}
There is a main method on Paper which adds new RealPlots to the
database. To produce new PredictedPlots, the appropriate
HZTool routines must be provided and called, and the Fortran
modified to output XML data files for the new histograms.

\subsection{To add a new simulation package}
The major work here would be to interface the new package to HZTool
(or an equivalent package capable of producing suitable output files - see
below). In addition, tables would have to be added to the database
specifying any extra parameters needed to define a model with the new
simulation package.

\subsection{To add information or new formatting to the web pages}
The ucl.hep.jetweb.html package and specifically the HTMLWriter class
is responsible for constructing the JetWeb web pages. Thus, changing
the layout or content of these pages necessitates modification of this
class's methods.

\subsection{To add servlet queries}
The user requests fits by filling in an HTML form. A
ucl.hep.jetweb.ResultSearchPattern and a ucl.hep.jetweb.PlotSelection
are built from the HTML form. A database query is constructed from
these objects. To add new parameters to user queries,
the relevant fields must be added to ResultSearchPattern and
PlotSelection, and the database query methods updated to incorporate
these new fields.

\subsection{To add HTML queries}
The prepared searches available from the welcome page are simple HTML
forms that are submitted to the server. These forms can be used as
templates for users to prepare their own standard (or often repeated) searches.

\subsection{To input data}

\subsubsection{XML}
\begin{figure}
\psfig{file=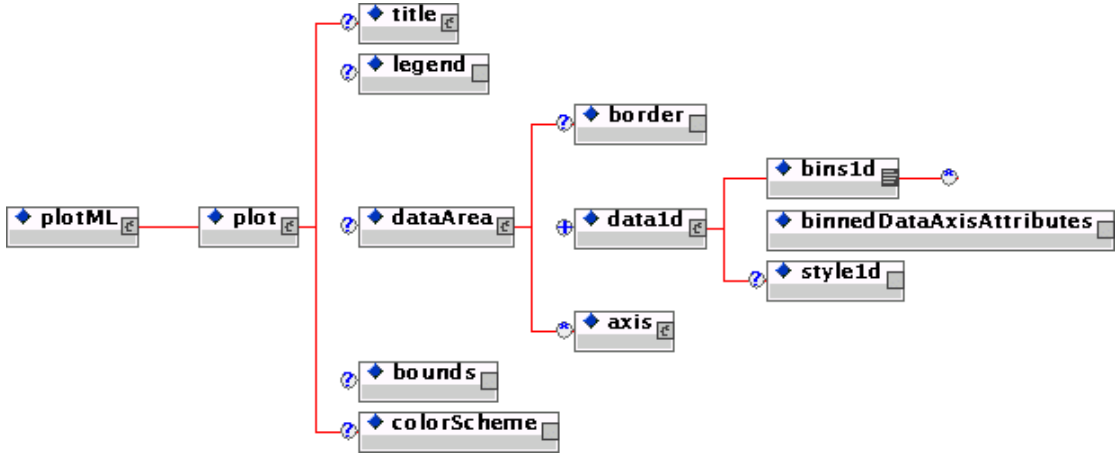,width=\textwidth}
\caption{\label{fig:plotMLfig} {\bf PlotML DTD - JetWeb reduced version}. This is a graphical representation of the tree-like format of
plotML. {\it plotML} is the root element of the document. Each reduced
plotML document contains a single {\it plot} element, made up of {\it title},
{\it legend},  {\it bounds}, {\it colorScheme} and the central {\it dataArea}. This {\it dataArea} element has a {\it  border}, an {\it axis} element and one or many {\it data1d} elements
(see Figure~9). This represents a major simplification of the original
plotML which, for example, also includes 2-dimensional histograms and
scatter plots.}
\end{figure}

Data is loaded into the JetWeb application in the form of XML files
generated by the HZTool FORTRAN routine. The format of these files is
derived from the JAS DTD - plotML. Although the XML files
produced are valid plotML files, in fact only a subset of the elements
in plotML are used. Thus, the JetWeb XML format can also be
represented by a reduced size plotML (Figures~\ref{fig:plotMLfig} and
~\ref{fig:plotMLdatafig}). 
An example XML file using this reduced plotML DTD is included in
the appendix.
The
ucl.hep.jetweb.plots.JetWebXMLReader reads in XML documents (reduced
plotML format) using the JDOM utility~\cite{jdom}. JetWebXMLReader
will ignore any additional elements present in the files. The
XMLReader utility could also be adapted to read
\begin{itemize}
\item a wider range of plotML documents;
\item a different XML format.
\end{itemize}
However, it would be simpler to retain the original plotML Reader and
use an XSL stylesheet to convert between XML formats, if this should
be required.

\begin{figure}
\psfig{file=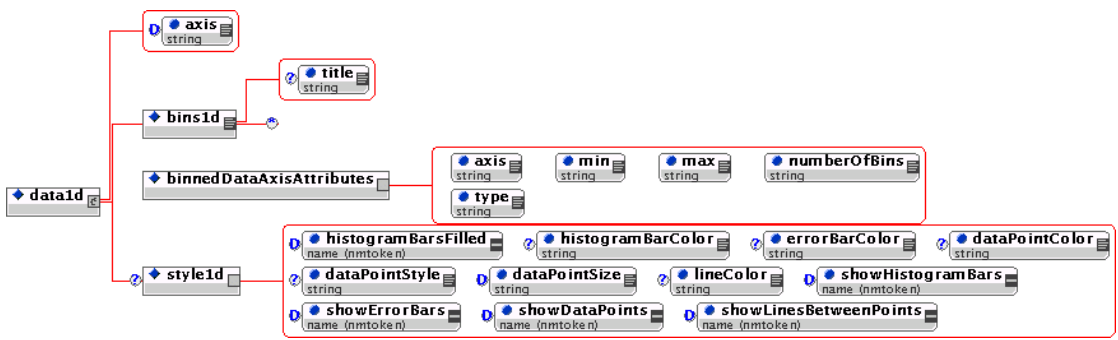,width=\textwidth}
\caption{\label{fig:plotMLdatafig} {\bf JetWeb 1D Histogram Data Element }. Agraphical representation of the {\it data1d} element of the jetweb reduced
version of plotML.The axes are specified within the
{\it binnedDataAxisAttributes}, while {\it styleId} specifies the appearance of
the plot. The data points themselves are held within the {\it bins1d}
element in a single (potentially very long) string, with each point
separated by a carriage return. For more detail, see appendix  and
the JAS documentation; ~\cite{plotML}. }
\end{figure}

One consequence of adopting the plotML input format for the java object
model is that any program capable of producing plotML histograms can be used
to provide simulation data for JetWeb. Currently a Fortran routine
(h2XML) is provided (and is available from the web pages) to
convert HBOOK histograms to the reduced plotML format described above. To
allow JetWeb to read data from other formats, similar conversion
facilities would be required.

\subsection{To output data}
\subsubsection{HTML}
The ucl.hep.jetweb.html package is responsible for producing the web
front end for JetWeb. Some of the classes (including Paper, DataPlot
and JobResult) have {\it toHTML} methods. These should really be
migrated to the HTML package in future releases.
\subsubsection{XML}
The JDOM functionality could be used in an XMLWriter class which would
reconstruct the XML documents from the object model. The documents
could be constructed in plotML, or another DTD.
\subsubsection{XML, HTML, any text format}
XSL stylesheets can be used to convert XML documents from one format
to another. If several formats were required, only a single XMLWriter
utility, used in conjunction with stylesheets, could produce multiple
output formats including XML, HTML - allowing direct visualisation of
data on a webpage - and plain text.
\subsubsection{Plots}
ucl.hep.jetweb.plots.DataPlot and its subclasses can be output as gif
files using the JAS utility. This utility is still under development
and supports an increasing variety of output formats, including
encapsulated postscript. If an unsupported format of plot file were
required, it would probably be simplest to extract the data in XML
format to be submitted to the plotting facility.

\subsection{Batch job submission}
Job submission from the server side is made in several ways. There are
options for submission to UK particle physics grid~\cite{GridPP} as
well as direct submission to local batch systems. This is transparent
to the general user. To submit to new facilties, template scripts must
be provided and the extra protocols added to the ucl.hep.jetweb.script
package.

\section{Future plans}
\label{sec:future}

There are several obvious and important developments of the JetWeb
facility which should take place. Some of these are described below.

Server side submission of jobs to the Grid has already been
implemented. It is also likely that the database itself will become
directly accessible as a grid data resource.

The Fortran library HZTOOL should eventually be augmented or replaced
by a similar package following an object-oriented design. This is
motivated by the advantages of OO design principals, the declining
knowledge base in Fortran and the likely future requirement to
validate C++ versions of the general purpose generators, PYTHIA7 \&
HERWIG++, which are currently under development. 

In addition to the new versions of those generators, several
next-to-leading order QCD generators are under development and these
could be tested with this facility.

Finally, there are major data sets and classes of data which
should be added to a truly `global' validation facility, for example LEP
and SLD hadronic data from the $Z$, HERA deep inelastic scattering data and
data from proton machines other than the Tevatron.

\section*{Acknowledgements}
Thanks to Brian Cox and Ben Waugh for useful discussions, encouragement
and CPU.

\section*{Appendix}
\subsection{HZTool XML output file - example}
\begin{verbatim}

<?xml version="1.0" encoding="ISO-8859-1" ?>
<!DOCTYPE plotML SYSTEM "plotML.dtd">
<plotML>
  <plot>
    <title>
      <border type="None"/>
      <label text="Jet ET &gt; 4 GeV">
        <font style="BOLD" points="14" face="SansSerif"/>
      </label>
    </title>
    <legend visible="never">
    </legend>
    <dataArea>
      <border type="None"/>
      <!-- 
	Each of the following data1d elements contains a single point.
	This handles variable bin widths.
	(For a constant bin width histogram, 
	 only one data1d element would be required,
	 with all data points in a single string, 
	 separated by carriage returns)
       -->
      <data1d axis="y0">
        <bins1d title=" ">
          0.49,0.14,0.14
        </bins1d>
        <binnedDataAxisAttributes max="-1.4" numberOfBins="1"
          min="-1.15" type="double" axis="x"/>
        <style1d dataPointSize="6" showErrorBars="true"
          histogramBarsFilled="true" showLinesBetweenPoints="false"
          errorBarColor="Black" dataPointStyle="dot" showHistogramBars="false"
          lineColor="Black" dataPointColor="Black" showDataPoints="true"
          histogramBarColor="Blue"/>
      </data1d>
      <data1d axis="y0">
        <bins1d title=" ">
          0.88,0.22,0.22
        </bins1d>
        <binnedDataAxisAttributes axis="x" numberOfBins="1"
          min="-0.8999999999999999" type="double" max="-1.15"/>
        <style1d dataPointSize="6" showErrorBars="true"
          histogramBarsFilled="true" showLinesBetweenPoints="false" 
          dataPointStyle="dot"
          errorBarColor="Black" showHistogramBars="false" lineColor="Black"
          dataPointColor="Black" showDataPoints="true" 
	  histogramBarColor="Red"/>
      </data1d>
      <data1d axis="y0">
        <bins1d title=" ">
          0.92,0.24,0.24
        </bins1d>
        <binnedDataAxisAttributes axis="x" numberOfBins="1"
          min="-0.7" type="double" max="-0.9000000000000001"/>
        <style1d dataPointSize="6" showErrorBars="true"
          histogramBarsFilled="true" showLinesBetweenPoints="false" 
	dataPointStyle="dot"
          errorBarColor="Black" showHistogramBars="false" lineColor="Black"
          dataPointColor="Black" showDataPoints="true" 
	  histogramBarColor="Fuchsia"/>
      </data1d>
      <data1d axis="y0">
        <bins1d title=" ">
          0.83,0.14,0.14
        </bins1d>
        <binnedDataAxisAttributes axis="x" numberOfBins="1"
          min="-0.5" type="double" max="-0.7"/>
        <style1d dataPointSize="6" showErrorBars="true"
          histogramBarsFilled="true" showLinesBetweenPoints="false" 
	  dataPointStyle="dot"
          errorBarColor="Black" showHistogramBars="false" lineColor="Black"
          dataPointColor="Black" showDataPoints="true" 
          histogramBarColor="Yellow"/>
      </data1d>
      <data1d axis="y0">
        <bins1d title=" ">
          0.85,0.17,0.17
        </bins1d>
        <binnedDataAxisAttributes axis="x" numberOfBins="1"
          min="-0.30000000000000004" type="double" max="-0.5"/>
        <style1d dataPointSize="6" showErrorBars="true"
          histogramBarsFilled="true" showLinesBetweenPoints="false" 
	  dataPointStyle="dot"
          errorBarColor="Black" showHistogramBars="false" lineColor="Black"
          dataPointColor="Black" showDataPoints="true" 
          histogramBarColor="Lime"/>
      </data1d>
      <data1d axis="y0">
        <bins1d title=" ">
          0.78,0.14,0.14
        </bins1d>
        <binnedDataAxisAttributes axis="x" numberOfBins="1"
          min="-0.10000000000000003" type="double" max="-0.3"/>
        <style1d dataPointSize="6" showErrorBars="true"
          histogramBarsFilled="true" showLinesBetweenPoints="false" 
	  dataPointStyle="dot"
          errorBarColor="Black" showHistogramBars="false" lineColor="Black"
          dataPointColor="Black" showDataPoints="true" 
          histogramBarColor="Orange"/>
      </data1d>
      <data1d axis="y0">
        <bins1d title=" ">
          0.54,0.12,0.12
        </bins1d>
        <binnedDataAxisAttributes axis="x" numberOfBins="1"
          min="0.04000000000000001" type="double" max="-0.1"/>
        <style1d dataPointSize="6" showErrorBars="true"
          histogramBarsFilled="true" showLinesBetweenPoints="false" 
	  dataPointStyle="dot"
          errorBarColor="Black" showHistogramBars="false" lineColor="Black"
          dataPointColor="Black" showDataPoints="true" 
          histogramBarColor="Aqua"/>
      </data1d>
      <axis logarithmic="false" showOverflows="false"
        numberOfBins="50" allowSuppressedZero="true" type="double" 
        location="x0">
        <label text="xgamma">
          <font style="PLAIN" face="Dialog" points="12"/>
        </label>
        <font style="PLAIN" face="Dialog" points="12"/>
      </axis>
      <axis logarithmic="false" showOverflows="false" min="0.0"
        allowSuppressedZero="true" type="double" max="1.0120000000000002" 
	location="y0">
        <label text="d(sigma)/d(xgamma) (nb)">
          <font style="PLAIN" face="Dialog" points="12"/>
        </label>
        <font style="PLAIN" face="Dialog" points="12"/>
      </axis>
    </dataArea>
    <bounds width="400" height="400"/>
    <colorScheme foregroundColor="default"
      backgroundColor="(255,215,0)"/>
  </plot>
</plotML>
\end{verbatim}

\end{document}